\documentclass{ws-procs9x6-cpt22}
\begin{document}

\newcommand{\refeq}[1]{(\ref{#1})}
\def\etal {{\it et al.}}
%any other macros go here 

\title{ Quark- and hadronic-level CPT violation phenomenology of neutral-mesons at at $d=5$ of the SME}

\author{A.\ Roberts,$^1$}

\address{$^1$ Indiana University Center for Spacetime Symmetries,\\
Bloomington, Indiana 47405, USA}

\begin{abstract}
This article summarizes an investigation of the connection between the hadron-level scalar effective field theory treatment of Lorentz symmetry violation in neutral-meson oscillations and the corresponding quark level $d=5$ nonminimal SME approach. The correspondence between the quark-level $a^{(5)}$ term is identified to first order in CPT violation (CPTV) based on the expectation that C, P, T symmetries and the spontaneously broken chiral symmetry of QCD should hold for both formalism.  
These can be related to the phenomenological parameter describing possible CPTV in the meson oscillations. A general result facilitating searches for kaons is given.  
\end{abstract}

\bodymatter

\section{Introduction}

The neutral-meson system revealed itself early on as a way to study CPT violation during the process of particle-antiparticle oscillations. These oscillations are described in the form of a Schrödinger-like equation in a Wigner-Weisskopf approximation yielding a 2$\times$2 effective hamiltonian for a two-meson system. Phenomenology and experiments in CPT symmetry studies are based on parametrizing differences between the diagonal elements of this effective hamiltonian pertaining to mass and decay rate differences of particles and antiparticles as mesons oscillate into one another. 

These studies have gone through considerable development as two schools of investigations grew. The Standard-Model Extension considered here is a general effective quantum field theory approach where CPT violation implies Lorentz violation and a momentum dependence is manifest in the phenomenology.\cite{KyCol} Testing of CPT in neutral mesons in this framework must be done depending on boost and orientation and a full phenomenological treatment must allow for flavor-dependent coupling to a Lorentz-violating background.\cite{NMPhKy}

The relevant terms have been shown to be those which match the C, P, T behavior of the hamiltonian of the mesons, where the violation is really a C violation pointing to the CPT-odd vector SME term
\begin{equation}
		\label{amu}
		{\cal L}\supset -\bar{q}\,a_q^{\mu}\gamma_{\mu} q\,,
	\end{equation} 
shown here for simplicity as the $d=3$ minimal SME term, with $q$ a quark field and $a_q^{\mu}$ a Lorentz-violating CPT-odd vector coefficient accounting for a coupling to a general Lorentz-violating background field causing the spacetime symmetry violation. This approach has been well tested in all neutral meson systems for the minimal renormalizable SME (mSME).\cite{datatables}
 
A further subtlety both theoretically and experimentally is to expand searches for the nonminimal fermion SME sector\cite{nonmin} and to investigate a deeper connection between a scalar effective field theory of the hadrons and the SME field theory formulated at the quark level as a lagrangian term of elementary particles. The fundamental parameters of the SME theory are the coefficients of these latter term. The phenomenological parameter is the one defined in the effective hamiltonian of the meson oscillations and must now include, besides the momentum dependence, the results of the quark-specific analysis. 

A first connection between the hadron- and quark-level treatment can be made to first order in CPTV using chiral perturbation theory and matching the discrete and gauge symmetries of the two descriptions.\cite{ChPT2} The connection to phenomenology must again include the proper analysis regarding the orientation and boost of each experiment. This talk presented this for the $a^{(5)}$ nonminimal coefficient for u,d, and s quarks in a chiral limit.

\section{Connection between the hadron- and quark-level theory}
The general form of the $a^{(5)}$ term for the fermion sector of the nonminimal SME was listed in the comprehensive work of Ref.\ \refcite{KoLi} as
\begin{equation}
	\label{a5def}
	{\cal L}_{\rm SME}\supset-\tfrac{1}{2}a^{(5)}{}^{\mu\alpha\beta}\bar{\psi}\gamma_{\mu}iD_{(\alpha}iD_{\beta)}\psi+\textrm{h.c.},
\end{equation}	
with $\psi$ denoting fermion fields. The SME preserves the Standard Model gauge symmetries and is included in the covariant derivative, and the Dirac gamma matrix is as usual. The spacetime indices of the covariant derivatives are symmetrized, indicated by the parentheses.  
The $d=5$ term in the scalar effective field theory in turn is
\begin{equation}
		\label{all_d_CPTV}
		{\cal L}_{\rm sEFT}\supset\frac{1}{2}\phi^\dagger(k_a^{(5)})^{\mu\alpha_1\alpha_2}\partial_\mu\partial_{\alpha_1}\partial_{\alpha_2}\phi+\textrm{h.c.}\,,
\end{equation}
with the ($k_a^{(5)})^{\mu\alpha_1\alpha_2}$ coefficient real, traceless, and symmetric in all Lorentz indices.\cite{EdKy19}
After proper symmetrization of the $a^{(5)}$ terms above, one finds that the discrete symmetries indeed match between free propagating $a{(5)}$ with the $k_a^{(5)}$.

To use the symmetries in chiral perturbation theory to connect mesons to quarks, one needs to go through a series of steps. This includes implementing left-handed doublets and right-handed singlets  into the lagrangian (\ref{a5def}), carrying the weak structure. Taken after that, the chiral limit reduces to an approximate $SU(3)_f$ structure for left and right-handed massless quark field vectors $\Psi_L\equiv(u_L,d_L,s_L)^T$ and $\Psi_R\equiv(u_R,d_R,s_R)^T$. In $u,d,s$ flavor space, two 3$\times$3 matrices$(a_L)^{\mu\alpha\beta}$ and $(a_R)^{\mu\alpha\beta}$ are introduced assuming the form 
\begin{align}
	\label{udsLR}
		\delta{\cal L}^{a^{(5)}}_{\rm qSME}=
		&{}-\bar{\Psi}_L(a_L)^{\mu\alpha\beta}\gamma_{\mu}iD_{(\alpha}iD_{\beta)}\Psi_L\nonumber\\
		&{}-\bar{\Psi}_R(a_R)^{\mu\alpha\beta}\gamma_{\mu}iD_{(\alpha}iD_{\beta)}\Psi_R\,.
	\end{align}
This is reduced appropriately to zero matrix entries where charge conservation mandates it. These chiral terms must later be recombined to give a quark-level lagrangian term of definite parity and charge conjugation. 

The goal is then to match the symmetries the SU(3)$\times$SU(3) symmetries for the meson and light-quark lagrangians. Following Ref.\ \refcite{XPT}, the mesons are identified as Nambu-Goldstone bosons of spontaneous chiral symmetry breaking, which mediate processes in the hadron approach in a low-energy context, as opposed to gluons at the quark level. 

These NG bosons are represented as
\begin{equation}
\label{Udef}
U(x)\equiv\exp\Bigg[i\dfrac{\lambda_a\phi_a(x)}{F_0}\Bigg]\,,
\end{equation}
where the $\lambda_a$ Gell-Mann matrices come from the SU(3) structure of QCD, $F_0$ is the pseudoscalar decay constant in the chiral limit,
and the $\phi_a$ correspond to the pseudoscalar meson octet. 

$U(x)$ transforms as  
$U(x)\to R U(x) L^{\dagger}$ under chiral transformations, where $(L,R)\in$ SU(3)$_L\times$SU(3)$_R$. The goal is then to build the lowest-order meson Lagrangian ($\delta{\cal L}_{\phi}$) from $U$
such that it carries the same quark transformation properties under C, P, T, and SU(3)$_L\times$SU(3)$_R$. 

This lagrangian contains left- and right-handed flavor-trace pieces and flavor octets from the quark coefficient matrices, $F_0$, an appropriate term of the $U(x)$ matrix above and some real-valued dimensionless constants. Restricting to free propagation and using the right decomposition of the quark coefficient matrices for the definite C, P, and T properties of the lagrangian will reduce the complexity of $\delta{\cal L}_{\phi}$ to 
\begin{align}
	\label{octet_L}
	\delta{\cal L}_{\phi} =
	& +i\frac{1}{2}\beta_8\, \phi_{\rm N}^{\dagger}\,(k^{(5)}_{\rm N})^{\mu\alpha\beta}\partial_\mu\partial_\alpha\partial_\beta\,\phi_{\rm N}+{\rm h.c.}\nonumber\\
	& +i\frac{1}{2}\beta_8\, \phi_{\rm C}^{\dagger}\,(k^{(5)}_{\rm C})^{\mu\alpha\beta}\partial_\mu\partial_\alpha\partial_\beta\,\phi_{\rm C}+{\rm h.c.},
\end{align}
with only one dimensionless undetermined parameter left. The neutral- and charged-meson fields have been separated into column vectors  
$\phi_{\rm N}\equiv(\pi^0,\eta,K_0)^T$ and $\phi_{\rm C}\equiv(K^+,\pi^+)^T$. These are not mixed electromagnetically. 
As an example, a result for $K_0$ can be taken. Some of the results can be further extrapolated to more massive quarks but that would require some suitable other constants covering the mass effects. 

Focusing on $k^{(5)}_{\rm N})^{\mu\alpha\beta}\partial_\mu\partial_\alpha\partial_\beta$, we arrive at the relationship set as the goal of the investigation:
\begin{equation}
\label{kdef}
(k^{(5)}_a)^{\mu\alpha\beta}_{K^0}=
\beta_8 \sum_{X=D,Q} \big[(a_X^{(5)}){}_{22}-(a_X^{(5)}){}_{11}\big]^{\mu\alpha\beta},
\end{equation}
where D and Q refer to components of a submatrix of the $(a_{L/R})^{\mu\alpha\beta}$ coefficients pertaining to the "d" and "s" quarks. 

What remains is to do a detailed analysis of the momentum-dependent phenomenology with explicit functions to describe detector geometry and orientation, sidereal harmonics, and boost dependence to test the $a^{(5)}$ coefficients component by component. The experimental analysis would, of course, also yield results for the coefficients of the scalar field theory, and has been analyzed in Ref.\ \refcite{EdKy19}.

\end{document}